\documentclass{egpubl}
\usepackage{EGUKCGVC2024}
 
\WsPaper           

\usepackage[T1]{fontenc}
\usepackage{dfadobe}  

\usepackage{cite}  
\BibtexOrBiblatex
\electronicVersion
\PrintedOrElectronic
\ifpdf \usepackage[pdftex]{graphicx} \pdfcompresslevel=9
\else \usepackage[dvips]{graphicx} \fi

\usepackage{egweblnk} 

\title[LLM-Assisted Visual Analytics]%
      {LLM-Assisted Visual Analytics: Opportunities and Challenges}

\author[M. Hutchinson, R. Jianu, A. Slingsby \& P. Madhyastha]
{\parbox{\textwidth}{\centering M. Hutchinson,
        R. Jianu,
        A. Slingsby
        and P. Madhyastha
        }
        \\
{\parbox{\textwidth}{\centering Department of Computer Science, City University of London
       }
}
}

\begin{document}


\maketitle
\begin{abstract}
We explore the integration of large language models (LLMs) into visual analytics (VA) systems to transform their capabilities through intuitive natural language interactions. We survey current research directions in this emerging field, examining how LLMs are integrated into data management, language interaction, visualisation generation, and language generation processes. We highlight the new possibilities that LLMs bring to VA, especially how they can change VA processes beyond the usual use cases. We especially highlight building new visualisation-language models, allowing access of a breadth of domain knowledge, multimodal interaction, and opportunities with guidance. Finally, we carefully consider the prominent challenges of using current LLMs in VA tasks. Our discussions in this paper aim to guide future researchers working on LLM-assisted VA systems and help them navigate common obstacles when developing these systems.

\ccsdesc[500]{Human-centered computing~Visual analytics}
\ccsdesc[300]{Computing methodologies~Natural language processing}
\printccsdesc 
\end{abstract}  
\section{Introduction}


Visual Analytics (VA) emphasises an analytical partnership between the computer and the human analyst, combining computational methods with interactive visualisation in an iterative process. However, effectively leveraging such systems requires expertise across analytical techniques, data visualisation principles, and domain-specific knowledge. This creates a high barrier to entry which can put powerful VA tools out of reach for many users. Moreover, analysing large, multi-faceted datasets typically involves iterative processes where visualisations and computational analyses have to be repeatedly configured and refined to probe new perspectives, hypotheses, and insights. The interactions needed to make this happen can often add significant overhead to the analysis process.



The advent of Large Language Models (LLMs) presents an increasingly viable solution to alleviate these limitations and support analysts within VA systems. By leveraging their vast knowledge bases and natural language processing capabilities, LLMs facilitate more natural, expressive and human-like communication, elevating VA systems to partners in the data analysis process. LLMs offer the potential to significantly lower barriers to entry for VA systems and streamline access to complex visualisations and powerful analytical tools.

Recent work has explored the intersection of VA and Natural Language Processing (NLP),  recognising the potential benefits of integrating language-based interfaces with visualisation tools. Shen et al. \cite{shen_towards_2023} review Visualisation-Oriented Natural Language Interfaces (V-NLIs), systems that support natural language (NL) input to produce visualisations. Similarly, Voigt et al. \cite{voigt_why_2022} focus on the use of NL in visualisation, covering systems that use NL either as an input or output modality. Wang et al. \cite{wang_survey_2022} and Wu et al. \cite{wu_ai4vis_2022} explored the use of machine learning and AI techniques in visualisation. 


This paper contributes to the ongoing discussion by exploring the integration of LLMs into VA systems, focusing on the opportunities and challenges associated with leveraging these powerful language models to enhance mixed-initiative VA systems. We explore how LLMs and their precursors have been used to support the VA pipeline (Section 3), discuss the opportunities and limitations of applying LLMs further in VA (Section 4), and finally conclude with a brief summary of the challenges associated with LLMs (Section 5). 


\section{Background}

VA systems have evolved significantly over the past decade, shifting from passive tools to active participants in the analytical process.
To facilitate this transformation, researchers have explored the development of mixed-initiative and adaptive VA systems. Mixed-initiative systems \cite{horvitz_principles_1999} are characterized by collaborative interaction between the user and the system, where both parties actively contribute towards a common analytical goal.  Adaptive systems are designed to continuously update their knowledge and behavior throughout the analysis process. This adaptivity allows the system to respond to the user's actions, preferences, and evolving understanding of the data.
Guidance aims to provide intelligent assistance to users throughout the analysis process \cite{ceneda_characterizing_2017}. Initially, only system-to-user guidance was considered, but later work acknowledged that guidance can be a mixed-initiative process, including guidance from the user to the system \cite{ceneda_review_2019}. Sperrle et al. \cite{sperrle_co-adaptive_2021} further developed this idea, proposing a model of co-adaptive guidance that emphasises the continuous adaptation of both user and system behavior and knowledge throughout the VA process as they guide each other.

In parallel, recent research has explored the integration of multimodal interactions to create shared analysis-contexts between users and VA systems. Multimodal systems leverage multiple input and output modalities, such as touch, gesture, gaze, and natural language, to enable more natural and intuitive interaction between users and the system \cite{lee_beyond_2012}. By providing users with a variety of interaction modalities, these systems aim to create a richer and more engaging analytical experience, facilitating the exchange of knowledge between the user and the system.

Building on the concept of richer interaction, the use of NL as an input modality in VA systems has also seen significant research. Early work on integrating NL into VA systems focused on using classical NLP pipelines, which often struggled with the ambiguity and noise inherent in human language. However, the emergence of LLMs has revolutionised the field of NLP, offering a more robust and flexible approach to language understanding and generation. LLMs, with their vast knowledge bases and advanced language processing capabilities, have the potential to not only interpret user input more effectively but also to generate human-like responses and assist in the analysis process. This paradigm shift is opening up new possibilities for leveraging NL in VA systems, moving beyond simple query interpretation to more complex tasks.

\section{Existing Systems}


VA has recently explored the use of LLMs to support parts of the analytical process. We discuss the state-of-the-art of the use of LLMs in VA over four emerging themes: data management, language interaction, visualisation generation, and language generation. Using representative works, we illustrate how LLMs have been used so far to enhance VA systems in each of these areas, discuss the efforts' limitations, and how they can be driven further. 


\subsection{Data Management}


Data lies at the core of any analytical process, and the effective management and preprocessing of data are crucial for deriving meaningful insights. Traditional VA systems often require users to provide pre-processed data in a specific format, such as tabular or other structured data formats, before analysis can commence. The integration of LLMs into VA systems provides opportunities for handling diverse data types, including unstructured data sources, alleviating the burden on users and expanding the range of data sources that can be leveraged for analysis. 


LLMs have demonstrated capabilities in processing and understanding unstructured text data. Zhang et al. \cite{zhang_large_2023} evaluate the performance of various LLMs across four data preprocessing tasks: error detection, data imputation, schema matching, and entity matching. GPT-4, in particular, gets good results, showing that LLMs can be used as effective pre-processors. The NL2Rigel \cite{huang_interactive_2023} system leverages GPT-3.5 to produce tabular data from raw data and NL user input. It also illustrates the efficacy of LLMs in transforming unstructured data into tabular data, and is even able to construct tables from webpages. 

LLMs have also shown the ability to generate data. Borisov et al. \cite{borisov_language_2023} demonstrate that LLMs can generate realistic synthetic tabular data across domains. Tavast et al. \cite{tavast_language_2022} show that LLMs can generate human-like emotion data, a useful capability in situations where the quality of the dataset, such as noise or missing values, could impede analysis. Synthetic data generation could also be a valuable tool for developing and testing visualisation concepts in the absence of real data. 

LLM-based systems also show promise in helping users retrieve relevant data for their analysis. Data-Copilot \cite{zhang_data-copilot_2024} was designed for data-related tasks: it first creates a set of versatile tools by analysing the available data and \textit{anticipating} user needs; then, when a user makes a request, Data-Copilot selects and uses the appropriate pre-designed tools to retrieve, process, and present the relevant data. 

This approach highlights the potential for LLMs to assist in not only the processing of given data but the collection and assembly of new relevant data from external sources (e.g., structured and unstructured data found online and parsed into usable formats). Data-Copilot can generate multi-form outputs such as text, tables, and visualisations, demonstrating the capability of future LLM use to streamline the data collection, analysis, and visualisation process in an end-to-end manner. This has the potential to allow analysts to focus on higher-level analytical reasoning, while the intuitive natural language interfaces make data exploration accessible to even non-technical users.

It is important to acknowledge the limitations associated with these approaches. One major concern is the potential for LLMs to introduce biases or inaccuracies based on their training data, impacting the data quality. The lack of transparency in how LLMs process and generate data can make it difficult for users to assess the reliability and provenance of the information. The use of LLMs in data management may also raise privacy concerns, as the data used to train these models might extend beyond publicly shared information. As LLMs are primarily text-based, there may also be limitations in their ability to handle complex data types, such as geospatial or time-series data, which are common in VA tasks. Another challenge lies in seamlessly integrating these LLM-based data management techniques into the VA workflow, as they are mostly separate systems currently.

\subsection{Language Interaction}

There has been significant research using NL as an input modality for VA systems, providing users with a more intuitive and accessible means of interaction. However, systems using rule-based NLP have faced challenges in handling the complexity and ambiguity of human language. 

Recent research in VA have explored various system types that incorporate NL capabilities to different degrees. One category is Visualisation-Oriented Natural Language Interfaces (V-NLIs) that given data, directly go from NL queries to generating corresponding visualisations \cite[interalia]{gao_datatone_2015, setlur_eviza_2016, narechania_nl4dv_2021}. Another related area is  that of visualisation recommendation systems and search interfaces that allow users to input NL queries and output ranked recommendations \cite[interalia]{luo_deepeye_2018_2, oppermann_vizcommender_2021}. Visual Question Answering systems form another category, enabling users to ask questions about charts \cite[interalia]{song_gvqa_2023, kim_answering_2020}. These systems are typically categorised based on the specific tasks they are designed to handle, as rule-based systems require customised development for each task. However, the emergence of LLMs offers the potential to unify these systems, as they possess capabilities across multiple tasks.

One significant challenge users face when interacting with rule-based natural language systems is the "cold start" problem. This issue arises when users struggle to initiate or continue an analysis due to a lack of understanding of the types of queries supported by the system \cite{srinivasan_discovering_2019}. This is because rule-based systems can only understand a limited set of instructions, and often struggle to handle underspecified or ambiguous language. LLMs have the potential to alleviate these problems as they have been demonstrated to perform at human level at natural language understanding related tasks. LLMs are often trained across multiple languages, so they can support input languages beyond English. As they can interpret a wider range of queries, users can express themselves more intuitively. This reduces the cognitive burden on users, enabling them to focus on their analytical goals rather than the specific capabilities of a given system. For example, Chat2Vis \cite{maddigan_chat2vis_2023} and ChartGPT \cite{tian_chartgpt_2024} demonstrate use of different GPT models to generate data visualisations, and users can express themselves in natural language without having to adhere to system rules. Some LLM-powered systems still incorporate prompt guidance: LIDA \cite{dibia_lida_2023} has a goal explorer to help users identify potential analytical goals, which can be useful assistance for novice users who may struggle to identify what they want to discover from the data.

Prior to the advent of LLMs, several systems have supported iterative NL interaction, capturing interaction history context to support analysis \cite[interalia]{lee_boomerang_2021, mitra_facilitating_2022}. These rule-based systems rely on a pipeline of models. Each of these components can introduce errors and uncertainties, especially when run iteratively. In contrast, LLMs can directly interact with human feedback and inputs, eliminating the need for complex pipelines and reducing the potential for errors.

\subsection{Visualisation Generation}

V-NLI systems have introduced the automatic generation of visualisations from user NL,  shifting the task of creating a visualisation specification from the user to the system. Even with the advent of LLMs, specification languages are still used to define visualisations. We discuss the use of LLMs to generate visualisation specifications, and highlight the application of image-based generative techniques in modifying visualisations.

Visualisation grammars are a well-established tool in VA, allowing users to define and encode visual mappings for their data. These languages, such as Vega-Lite \cite{satyanarayan_vega-lite_2017}, provide a structured and declarative way to describe the desired visual representation, including the data transformations, encoding channels, and layout properties. Many existing V-NLI systems rely on these specification languages to generate visualisations based on user inputs. For example, NL4DV generates vega-lite specifications \cite{narechania_nl4dv_2021, mitra_facilitating_2022}.

Visualisation specification languages continue to play a crucial role as a bridge between NL inputs and the actual rendering of visualisations. For example, Chat2Vis \cite{maddigan_chat2vis_2023} demonstrates GPT-3, Codex, and ChatGPT to generate visualisations from user queries, first generating Python code based on the query which is then used to produce the visualisation. LIDA \cite{dibia_lida_2023} works similarly, generating python code to produce visualisations from user queries. ChartGPT \cite{tian_chartgpt_2024} produces Vega-Lite specifications from user queries in a structured way, generating answers to sub-tasks defining the filter, mark, encoding, and sort and combining them to form a visualisation.


There are also image-based generative techniques that have been applied to the design aspects of visualisation. LIDA \cite{dibia_lida_2023} uses a text-to-image generation model to turn visualisations into stylised infographics based on user prompts. ChartSpark \cite{xiao_let_2024} employs a similar approach to generate pictorial visualisations. These systems demonstrate the potential for leveraging generative models to create visually appealing artistic representations of data. Schetinger et al. \cite{schetinger_doom_2023} offer a comprehensive review of previous work and opportunities for text-to-image generative models in data visualisation.

\subsection{Language Generation}

NL has been used as an output modality to communicate insights, findings, and explanations to users in VA systems. Rule-based systems are limited in the flexibility and diversity of the text that they can produce. LLMs have capabilities in Natural Language Generation (NLG), producing human-like text, meaning they can overcome the limitations of previous systems. We explore how NL has been used to communicate with users in VA systems, discussing how the flexibility of LLMs can address the shortcomings of previous approaches, whilst also acknowledging the limitations of the current applications of LLMs in this context.

Before LLMs, early research exploited rule-based NLG techniques to communicate insights to users. Systems like Calliope \cite{shi_calliope_2021} and Voder \cite{srinivasan_augmenting_2019} produce data facts, NL descriptions of statistical facts about the data used. These systems typically rely on a limited set of predefined fact types and template-based generation methods, limiting the diversity of language that they can produce. Similarly, some systems generate captions or titles for visualisations using template-based approaches \cite{hsu_scicap_2021}.

LLMs have the capability to generate more flexible and diverse NL compared to rule-based systems. Some recent systems have leveraged LLMs to generate individual facts or annotations to supplement visualisations. For example, the InkSight \cite{lin_inksight_2023} system uses an LLM to generate annotations from user sketches on visualisations. However, this approach still relies on a template based approach. Due to the limitations of LLMs in analytical reasoning, the statistical facts about the data are generated separately and the LLM is only used to generate more fluent NL. LLMs have also been used to construct entire narratives. DATATALES \cite{sultanum_datatales_2023} is a prototype system that leverages an LLM to help users author data-driven articles based on a given chart and user annotations.

While these LLM-based approaches demonstrate the potential for more flexible and contextually relevant NL in VA systems, they still face limitations. LLMs struggle with analytical reasoning, and may generate text that is fluent, but not always accurate to the underlying data. For this reason, many LLM-based NLG approaches in VA still rely on templates, limiting their ability to fully leverage the flexibility of LLMs.

\section{Opportunities}

In this section, we explore several key areas where LLMs show promise in advancing the field of VA. We explore opportunities by considering the broader VA pipeline alongside gaps and open problems in the existing work. We cover data integration, visualisation generation, leveraging domain knowledge, multimodal interaction, and guidance.  By examining these opportunities, we want to highlight the potential for LLMs to significantly enhance the capabilities of VA systems, making them more powerful, intuitive, and adaptable to diverse user needs. The integration of LLMs could lead to the development of a new generation of VA tools that can better support users in their analytical tasks.

\subsection{Visualisation-Language Models}
Current visual analytics systems that incorporate natural language interactions often rely on code as a bridge between the user's input and the resulting visualisation. This approach, while effective, can limit the types of visualisations that can be generated, as the system is constrained by the expressiveness of the underlying programming language or library used.

There is opportunity to develop more flexible visualisation systems using LLMs. One possibility is to use a more versatile bridge, such as D3.js, which provides a wide range of capabilities for creating interactive visualisations compared to python libraries or Vega-Lite. 

Another prospect is to generate visualisations directly, without the need for an intermediary programming language at all. LLMs have demonstrated remarkable capabilities in understanding and generating text, but they rely solely on text-based input and output. Recent advancements in multimodal machine learning have led to the development of vision-language models that can understand and process visual information alongside text. These vision-language models are trained on large datasets of both text (usually captions or descriptions) and diverse naturalistic images. This enables them to learn the relationships and correspondences between images and text, capturing visual knowledge that is currently untapped \cite{li_multimodal_2023}.

Such models have been primarily applied to general computer vision tasks on natural images and creative applications. A promising direction for VA is to extend such models to reason effectively about visualisation-specific stimuli (visual data encodings) alongside their NL interpretations (e.g., descriptions of data shown or findings facilitated). We refer to these as \textit{Visualisation-Language models}. By enabling models to directly interpret visual representations, rather than relying on intermediate text-based specifications, multimodal models could offer a more intuitive and flexible approach to generating and manipulating visualisations, moving beyond the expressivity of particular specification languages. This could expand the range of possible visualisations and allow for more natural and expressive interactions between the user and the system.  While this approach is still in its early stages, it holds promise for creating highly customisable and diverse visualisations.

Moreover, Visualisation-Language models could also be leveraged to interpret and understand existing visualisations. Recent research has explored reverse engineering both visualisation specifications and the underlying data from bitmap images using pipeline and neural network-based techniques \cite{poco_reverse-engineering_2017, jung_chartsense_2017}. Training such models on research paper figures and illustrations in the news and popular media across a broad range of domains could harness a large, untapped body of knowledge to link data, visualisation design, and the insights they communicated or conclusions they supported. 

However, this direction also presents challenges, particularly in interpreting and generating interactivity, which is a crucial aspect of visual analytics. Generating static images may limit the user's ability to explore and interact with the data effectively. Additionally, the current state of reasoning capabilities in LLMs can make it difficult to specify exact requirements when generating images directly. LLMs may struggle to understand and incorporate complex constraints or design principles, leading to visualisations that may not effectively convey the intended information or insights.

\subsection{Domain Knowledge}
Domain-specific VA are typically custom-engineered to cater to a particular domain following design study methodologies \cite{shen_domain-specific_2021}. These systems are loaded with information about domain-specific conventions and procedures, often using their own domain-specific languages.  While this approach allows for more tailored and efficient analysis within a given domain, it also limits the flexibility and adaptability of these systems to other domains or use cases.

The emergence of LLMs has opened up new possibilities for incorporating vast amounts of domain knowledge into VA systems. LLMs are pretrained on massive corpora of text data spanning multiple domains, allowing them to capture and understand a wide range of domain-specific information \cite{bossmani_opportunities_2022}. This inherent domain knowledge can be leveraged to create more versatile and adaptive VA systems that can cater to various domains and user needs. A single LLM-powered VA system could potentially assist users in analysing any type of data without the need for separate, domain-specific implementations. Using this training data, LLMs could offer assistance with VA tasks based upon practices that are common within the domain. This type of assistance could help users align their analyses with well established practices and techniques, potentially improving the quality of their findings.

LLMs can be further fine-tuned to specific domains. By training LLMs on additional domain-specific datasets, or incorporating domain rules or constraints, their knowledge can be refined to better suit the needs of a specific domain. Further, methods such as chain-of-thought-prompting, that require very few detailed demonstrations, offer an efficient and effective means to seamlessly integrate domain specific expertise into LLMs\cite{wei_chain--thought_2023}.



However, there are limitations to these approaches using LLMs for domain specific applications. Whilst they are trained on a broad range of data, they may not actually have an understanding of domain specific techniques in practice, or generate information that is not actually true. As mentioned, this could, in some cases, be alleviated with further fine-tuning or by adding additional constraints. There is also a lag between the creation of new knowledge in a domain and the incorporation of that data into an LLM, as their data is usally collected at a certain point in time. This could lead to the generation of out-of-date information that may not actually reflect the best practices in a given domain. They are also not currently able to justify what particular data led them to make a specific recommendation, which poses a significant challenge for more specialised domains. More research is required to improve the reliability of LLMs in the context of domain-specific systems. 


\subsection{Multimodal Interaction}



The integration of LLMs into VA systems has led to a growing focus on conversational interfaces that rely primarily on natural language input. While this approach has shown promise in enabling more intuitive and accessible interactions, there is an opportunity to further enhance the user experience by combining language-based input with other well-established interaction modalities.

Direct manipulation techniques which underlie WIMP (Windows, Icons, Menus, Pointer) interfaces have been fundamental to supporting interaction in VA systems to date and are likely to continue playing a significant role due to their effectiveness in supporting certain tasks. For example, selecting data points, zooming into specific regions of a visualisation, or adjusting parameters through sliders can be more easily accomplished through direct manipulation than with language alone. By integrating NL alongside these traditional interaction modalities, VA systems can offer a more powerful and flexible user experience that combines the best of both interaction paradigms.

Researchers have already begun exploring the integration of various interaction modalities alongside natural language input in LLM-based systems. For instance, the InkSight system \cite{lin_inksight_2023} combines sketch-based input with natural language, allowing users to annotate visualisations and generate insights. Similarly, LLMs have demonstrated capabilities in speech recognition \cite{hu_large_2024} and gesture recognition \cite{wicke_probing_2024}, opening up new possibilities for integrating these modalities more widely into VA systems. By leveraging multiple input channels simultaneously, systems can gain a more comprehensive understanding of the user's intent and level of understanding throughout the analysis process.

However, there are also limitations to consider when integrating multiple modalities alongside NL. One challenge is the complexity of processing multiple input modalities in real-time. Systems would need to be able to capture and interpret several different input modalities simultaneously, whilst also handling any conflicts or ambiguities in these different channels. Systems may need specialised hardware to capture certain types of inputs, which may not be feasible in application. Further research is required to explore how to effectively implement LLM-based multimodal interaction techniques and their impact on user experience.

\subsection{Guidance}

Guidance is a key aspect of many VA systems. Current systems mostly follow user queries directly in a reactive manner. Ceneda’s model of guidance \cite{ceneda_characterizing_2017} includes three different levels: orienting, directing, and prescribing. With their ability to accurately understand and infer from  human language, LLMs could adapt the level of guidance provided based on the user's knowledge gap and the complexity of the task at hand, rather than just following user instruction.

In Ceneda's model, guidance in VA incorporates interaction history. LLMs have the potential to leverage this rich interaction history, including user queries and the dataset itself to provide guidance to users. This could involve proactively conducting analyses on the user's behalf and unobtrusively presenting relevant findings or insights that align with the user's objectives. 


Sperrle et al. \cite{sperrle_co-adaptive_2021} explore the concept of co-adaptive guidance in mixed-initiative VA systems, also incorporating guidance from the user to the system. With their ability to both generate and understand language, LLMs have the capability to facilitate this two-way guidance. Through asking questions, providing options, and eliciting user feedback, LLMs can learn from users and adapt their behavior accordingly. This bi-directional communication can lead to a more collaborative and adaptive visual analytics experience, where both the user and the system continuously learn from each other.
Systems need to strike the right balance between proactive guidance and user autonomy.
\section{Challenges}

While the integration of LLMs presents enormous opportunities, as highlighted in our previous section, their application in the field of VA faces several obstacles that span various aspects of the analytical process. In this section, we discuss some of the most significant risks and obstacles associated with the use of current state-of-the-art LLMs in VA and emphasize the necessity of rigorous research to address them.

\noindent
\textbf{VA expertise:} LLMs often lack explicit domain-specific knowledge regarding VA principles and established best practices. Consequently, LLMs may generate outputs that, while linguistically plausible, are ineffective or misaligned with the goals of visual analysis. To mitigate this, we need new techniques to integrate LLMs with relevant VA knowledge. Such techniques could include reinforcement learning from human feedback (RLHF) aligned with VA objectives which are directly provided by VA experts or analysts, integration of rule-based constraints, or fine-tuning on a corpus of domain specific visualisation knowledge or past VA interactions. 

\newpage
\noindent
\textbf{Explainability and interpretability:} The black-box nature of LLMs poses another significant challenge, the opaque inference processes of these models make it difficult for users to understand how the system arrived at a particular output, especially relating to analytical processes. This lack of transparency hinders users' ability to understand how the system derives recommendations, potentially undermining trust and utility of the outputs. Interpretability and explainability of the system is particularly crucial in VA, where analysis accountability often directly impacts decision-making.

\noindent
\textbf{Evaluation:} Evaluating the effectiveness and usability of LLM-based VA systems requires a more comprehensive approach. Traditional metrics, such as those which are directly borrowed from NLP, may especially fail to fully capture the intricate interactions between the LLM, analyses, visualisation outputs, and user interaction. Developing comprehensive evaluation frameworks is essential; these frameworks should assess, for e.g., guidance quality, system adaptability across domains and tasks, and overall user experience. 

\noindent
\textbf{Data faithfulness:} Ensuring data-faithfulness, or the ability of LLMs to generate outputs that are consistent with the underlying data and domain knowledge, is a significant concern. LLMs have a propensity to `hallucinate’ or generate plausible but incorrect information, which can lead to flawed insights and decision-making\cite{mahowald2023dissociating}. Mitigating this risk requires the development of techniques to explicitly ground LLMs in the provided data and incorporate VA specific knowledge from reliable sources. 

\noindent
\textbf{Reasoning:}  As highlighted in Section 4, current LLMs exhibit limited abilities for complex reasoning and accurate conclusions from analytical data \cite{mccoy_embers_2023}. As VA is fundamentally an analytical reasoning task, this limitation poses a significant barrier to their seamless integration. Research into methods for augmenting LLMs with structured reasoning (specifically analytical, numerical and mathematical) capabilities is critical for their applications in VA.

\noindent
\textbf{Attributions:} The lack of reliable attributions and the potential for using outdated information pose additional challenges. LLMs often struggle to provide specific attributions or sources for the information they produce, making it difficult to verify the provenance and reliability of the generated content. This is especially challenging when opening an LLM-assisted VA system to non-experts. 

Addressing these challenges will be crucial for the successful integration of LLMs in VA processes. 

\section{Conclusion}

In this paper, we began by exploring current approaches to integrate LLMs into various stages of the VA pipeline. We identified key opportunities where LLMs could significantly elevate and advance VA capabilities. Finally, we presented a discussion on potential risks and challenges of developing LLM-assisted visual analytics. We hope that our work will help to position and advance novel visual analytic systems that take advantage of the significant potential that LLMs offer while mitigating their risks, allowing for development of efficient, accessible and intelligent VA processes.


\bibliographystyle{eg-alpha} 
\bibliography{updated_bib}       


\end{document}